\documentclass[
twocolumn,
]{ceurart}

\usepackage{listings}
\begin{document}

\copyrightyear{2025}
\copyrightclause{Copyright for this paper by its authors.
  Use permitted under Creative Commons License Attribution 4.0
  International (CC BY 4.0).}

\conference{Normalize at the 19th ACM Recommender Systems Conference, 
  September 22 - 26, 2025, Prague, Czech Republic}

\title{Civic Ground Truth in News Recommenders: A Method for Public Value Scoring}

\author[1]{James Meese}[%
orcid=0000-0003-1910-6166,
email=james.meese@rmit.edu.au,
]
\address[1]{School of Media and Communication, RMIT University,
  Latrobe St, Melbourne, Australia}

\author[2]{Kyle Herbertson}[%
orcid=0009-0001-3764-6222,
email=kyle.herbertson@rmit.edu.au,
url=https://github.com/KalykeXIII,
]
\address[2]{School of Media and Communication, RMIT University,
  Latrobe St, Melbourne, Australia}


\begin{abstract}
Research in news recommendation systems (NRS) continues to explore the best ways to integrate normative goals such as editorial objectives and public service values into existing systems. Prior efforts have incorporated expert input or audience feedback to quantify these values, laying the groundwork for more civic-minded recommender systems. This paper contributes to that trajectory, introducing a method for embedding civic values into NRS through large-scale, structured audience evaluations. The proposed civic ground truth approach aims to generate value-based labels through a nationally representative survey that are generalisable across a wider news corpus, using automated metadata enrichment.
\end{abstract}


\maketitle

\section{Introduction}
Research in news recommendation systems (NRS) has recognized the importance of 'beyond precision' metrics, working to design recommender systems that support  normative outcomes. Most notably, Sanne Vrijenhoek and colleagues have established evaluation metrics \cite{vrijenhoek2024}, which can quantify the extent to which the output of recommendation systems aligns with normative standards. Two of these metrics have been empirically tested: activation, which refers to 'variations in journalistic writing style' and alternative voices, 'the extent to which minorities are being represented' \citep [p. 652] {mattis2025}. It was found that certain deployments had a limited effect on users with audiences showing 'limited appetite for extreme versions of news recommenders that adhere to ideals of critical models of democracy' \citep [p. 663] {mattis2025}. 

Alongside this scholarly activity, similar efforts are being championed in industry contexts. Expert feedback is a common mechanism, where editorial objectives are embedded in NRS. For example, Swedish Radio has developed a public service algorithm allowing journalists to signal whether their story addresses key public service goals through metadata labelling. The Dutch public service organisation NPO (Nederlandse Publieke Omroep) also has an active audience survey, polling thousands of audience members daily about the public value of content they have watched through a mobile app. This enables the public service media organisation to generate a public value metric, which they use as a weighting mechanism within their wider recommender system. 

These different industry and scholarly projects provide some useful insight into the current state of normative NRS. On the academic side there are theoretically-informed \cite{helberger2021} normative tools enabling post-hoc auditing to be conducted \cite{vrijenhoek2024, vrijenhoek2024diversity} and various empirical assessments are underway. In parallel, industry deployments are increasing. Our paper details efforts to advance this work by devising structured public value signals from a rating task, which can be extrapolated across a wider news corpus. Subsequent user evaluation will assist with further refinements before it is ultimately used as a weighting signal. The method is called civic ground truth and aims to build on the NPO model. Our paper proceeds as follows. We begin by outlining our concept of civic ground truth and setting out the contribution to the field, before detailing the methods for our first wave study, which involves metadata enrichment and a representative national survey, followed by semi-supervised extrapolation to generate public value scores across a corpus of news. We end with a brief discussion and conclusion. 

\section{Civic Ground Truth}
We define civic ground truth as audience-generated judgements about the societal importance of news. While the term ground truth implies a stable or coherent benchmark, socio-technical research has challenged these definitions noting the often provisional and ad-hoc nature of these datasets \cite{jaton2025examining}. Therefore, while the opinions of the public may fluctuate over time, we suggest they still provide a vital normative baseline for recommendation. In doing so the project directly accounts for the views of citizens, in contrast to current work where user input has been productively sought for empirical testing rather than value construction \cite{heitz2022, heitz2023deliberative, mattis2025, shin2024value}. 

We capture these views through classification tasks conducted as a nationally representative survey, where participants rate items based on perceived personal and public interest. Respondents are guided with a brief description of the public interest and are then asked to also rate items based on certain sub-dimensions of public value. As the literature has noted, the concept of the 'public interest' is 'broad, vague and loosely constructed' \citep [p.25] {iosifidis2011media} despite being a foundational concept of media law and policy \cite{napoli2019social, pickard2019democracy, christians2009normative}. However, recent work has shown that normative goals can be translated into technical areas despite these conceptual challenges \cite{vrijenhoek2024}. Our scale continues this work by drawing on legislative descriptions of what news is considered valuable to Australians \cite{newsmedia}, previous conceptual efforts to identify 'public affairs' content \cite{dwyer2023media} and pre-existing normative implementations such as RAdio \cite{vrijenhoek2024}. 

Such an approach means that our labelling process is not wholly bottom-up, as respondents are given normative frames to evaluate content against. However, we offer a path towards a representative evaluation of a media corpus, rather than just capturing ratings post-consumption as NPO does. We also use these evaluations to inform future recommender testbeds and develop civic weightings that can inform re-ranking algorithms. Our survey captures the civic orientation of respondents drawing on previously validated batteries \cite{prochazka2019measure} and our own scale, which will be validated through expert review, Exploratory Factor Analysis and Confirmatory Factor Analysis. These questions will enable us to capture a range of civic profiles from issue-specific engagement with news to respondents who are disengaged with news. We now outline these methods in more detail below. 
   
\section{Methodology}

A persistent challenge in NRS research is the lack of sustained access to high-quality news datasets \cite{vrijenhoek2023mindreflectionsminddataset}. While some datasets are available, they are typically released as static snapshots rather than actively maintained resources. This makes reproducibility and longitudinal analysis difficult. In the Australian context, such datasets are particularly scarce.

To address this limitation, this study adopts a dataset-agnostic approach by relying solely on information that is regularly and publicly available to news consumers. This includes article headlines, featured images, bylines (author names), and publication dates. This offers greater methodological adaptability to future data sources, irrespective of their origin.

Rather than sampling from existing datasets, the study constructs a 'complete' dataset, collecting all articles published on the aggregate news site \textit{News.com.au} over a one-month period. Securing a representative dataset of all news in Australia is challenging, especially given growing restrictions around web scraping. Given this wider context, we have selected one of the most visited commercial news websites in Australia as a mainstream and familiar source with a variety of news. The dataset is then pre-processed to normalise key metadata fields, including the article headline, author name, and publication date.

To support generalisation from the labelled data to a broader corpora, we enrich each article with a structured set of metadata. This is achieved using automated headline-level news value extraction techniques adapted from Piotrkowicz et al. \cite{wrro110978}. Adaptation of this methodology included transitioning to ``\verb|flair|'' and ``\verb|nltk|'' packages, otherwise remaining true to the evaluations described. A selection of these measures are detailed below.

While not comprehensive, this enrichment process was selected for consistency and scale, enabling this process to be reliably applied to articles in future datasets.

\subsection*{Prominence}
Prominence aims to capture the inherent visibility of a headline based on the recognisability and popularity of the entities it mentions. This is operationalised by measuring Wikipedia page view statistics for each entity in the headline \cite{arrendondo2023mediawiki, mediawiki_api_2025}. The approach generates statistics for both short-term and long-term prominence, as well as a boolean judgement of burst \cite{10.1145/1007568.1007586}. While this method tends to favour popular-culture content, or that related to celebrities, politicians, and other well-known figures, it offers a proxy for audience interest and reach in the absence of proprietary viewership data. As such, prominence is understood here as an externally observable signal of potential audience attention, grounded in collective online behaviour.

\subsection*{Sentiment}
Where prominence captures attention driven by entity recognition, sentiment focuses on the linguistic affective qualities of a headline. Sentiment analysis, an established technique in natural language processing, helps to assess how emotionally charged or polarising a headline may be, regardless of the prominence of the people or topics it mentions. Headlines lacking recognisable entities may still achieve widespread circulation due to provocative, emotionally engaging, or sensational wording. Thus, sentiment serves as a complementary measure to prominence, capturing forms of engagement that arise from language-based persuasion or affective appeal, rather than from recognition and popularity alone. 

Features relating to sentiment and emotionality have been shown to influence a news article’s virality \cite{berger2013}. While the predictive power of sentiment in the context of headlines specifically remains unproven (as noted in \textit{Automatic Extraction of News Values from Headline Text}) this study includes sentiment not as a predictive feature but as a descriptive metadata field. Its inclusion is intended to capture a relevant linguistic property of headlines that may interact meaningfully with other elements of news value, even in the absence of direct empirical validation.

\subsection*{Surprise}
In addition to prominence and sentiment, surprise captures another important linguistic feature of news headlines: how much a headline defies reader expectations through unusual or unexpected combinations of words. As defined in \textit{Automatic Extraction of News Values from Headline Text}, 'surprise' is understood here as explicit linguistic surprise; quantified via the statistical measures of phrase co-occurrences within the headline.

To extend this notion, our research also includes other linguistic features, such as measures of clickbait-likeness, based on the work of Chakraborty et al. \cite{DBLP:journals/corr/ChakrabortyPKG16}. These features look for common techniques used in engagement-driven headlines, such as curiosity gaps, lists, rankings, or second-person address.

\subsection*{Labelling}
Using these features, we apply k-means clustering to group articles based on shared thematic and stylistic characteristics derived from the enriched metadata. A stratified sample is then drawn from across these clusters to ensure coverage of the dataset’s diversity. Each respondent in a nationally representative survey panel will evaluate a subset of articles, randomly selected from this stratified sample, using a consistent set of questions aligned with normative media policy goals. The exact number of clusters and articles presented per respondent will be determined based on data distribution and pilot testing, with attention to balancing coverage and respondent burden.

Finally, by linking survey responses to both demographic variables and the enriched metadata, we create a model that maps civic orientations and values onto measurable article features. This model serves as a bottom-up recommendation mechanism that can be applied to future datasets, using the labelled subset as training and test data for predictive extrapolation.

\section{Discussion and Conclusion}

The above project maps a pathway for involving citizens in value construction and ranking processes, both developing and expanding on current industry applications and extant studies. It will not replace other valuable methods associated with normative recommendation, such as expert-informed labelling or recommender evaluation protocols \cite{vrijenhoek2024diversity}. However, a representative assessment of stories offers the potential for whole-of-society modelling of what publics perceive as civically valuable.  

Our method also supports the systematic identification of civic orientations, and development of plural scoring inputs for subsequent optimisation. It may even expand the range of recommendation options beyond existing reforms (like the DSA), which offer users a binary option of either remaining within personalised algorithmic systems or opting out. Instead, our approach could support re-ranking logics that align with multiple civic priorities based on the needs of different publics. 

The approach practically operationalises current legal scholarship that proposes the development of an individual right to ‘constructive optimisation’ of recommendation systems \cite{naudts2025right}. It also proposes a model of participatory recommender governance, allowing national publics to shape what is surfaced and prioritised. Importantly, the method recognises civic pluralism, ensuring that minority civic viewpoints are not overridden by majority opinions or expert judgement. 

There are limitations, the most prominent being that our approach is not directly portable to commonly used datasets available within the recsys community, such as G1 or MIND. As such while our approach offers insights as a proof of concept, it will not be usable as a benchmark or method of evaluation. At the project level, we also anticipate some respondents will be civically disengaged or disconnected from mainstream news content. For example, they might score certain news stories as important to society but not find them personally engaging. Alternatively, they may have civic orientation that is not aligned with mainstream news. As a result, working out what to recommend will involve further experimentation. Other more common limitations relate to standard challenges when dealing with surveys, from social desirability bias to priming effects. These challenges are unavoidable in representative survey work. In some cases, the may even directly support the project's aim, which is to frame citizen input in normative terms. 

This paper is our first attempt to set out this project, with the survey instruments and test corpus close to completion and survey and subsequent evaluation work to follow. Therefore much of what we propose is anticipatory in nature. However, we argue there are strong justifications for incorporating citizen feedback into normative recommendation during the re-ranking stage. While certain forms of exposure diversity have a limited impact on attitudes and behaviours \cite{mattis2025}, asking for representative civic input in rankings and a more nuanced weighting across different profiles may lead to increased satisfaction amongst civic-aligned users. There are also more normative justifications associated with including audiences in actual assessments of journalistic value, rather than making abstract claims, which is common in the news industry and amongst journalism scholars \cite{carlson2023journalism}. Given these potential benefits, we suggest that this approach offers one valuable way to progress the alignment of recommendation systems with the public good, and further support policy goals associated with media pluralism \cite{dwyer2023media}. 

\bibliography{references}

\section*{Declaration on Generative AI}
 During the preparation of this work, the author(s) used GPT-4o for: Paraphrase and rewording to improve writing style and Grammar and spelling check. After using these tool(s)/service(s), the author(s) reviewed and edited the content as needed and take(s) full responsibility for the publication’s content. 

\end{document}